\documentclass[10pt,preprint]{aastex}

\usepackage{graphicx}
\usepackage{times}
\usepackage{amssymb}
\newcommand{\prt}{\partial}
\newcommand{\mrm}{\mathrm}

\begin{document}

\title{Migration of a subsurface wavefield in reflection
seismics: A mathematical study}

\author{Arnab K. Ray}
\affil{Homi Bhabha Centre for Science Education \\
Tata Institute of Fundamental Research \\
V. N. Purav Marg, Mankhurd, Mumbai 400088, India}
\email{akr@hbcse.tifr.res.in}

\begin{abstract}
In this pedagogically motivated work, the process of migration in 
reflection seismics has been 
considered from a rigorously mathematical viewpoint.  
An inclined subsurface reflector with a constant dipping angle has been shown 
to cause a shift in the normal moveout equation, with the peak
of the moveout curve tracing an elliptic locus. Since any subsurface 
reflector actually has a non-uniform spatial variation, the use of 
a more comprehensive principle of migration, by adopting the wave 
equation, has been argued to be necessary. By this approach  
an expression has been derived for both the amplitude and the phase of 
a subsurface wavefield with vertical velocity
variation. This treatment has entailed the application of the 
{\rm WKB} approximation, whose self-consistency has been established by the
fact that the logarithmic variation of the velocity is very slow in the 
vertical direction, a feature that is much more strongly upheld at increasingly 
greater subsurface depths. Finally, it has been demonstrated that for a
planar subsurface wavefield, there is an equivalence 
between the constant velocity Stolt Migration algorithm and the 
stationary phase approximation method (by which the origin of 
the reflected subsurface signals is determined).
\end{abstract}


\section{Reflection seismology: A basic introduction}
\label{sec1}

In seismic processing, migration of seismic data is an exercise of paramount
importance. Unless this aspect of processing is addressed satisfactorily, 
the possibility of making misleading estimates of subsurface depths remains
wide open. Even by a simple quantitative estimate it can be shown that in many 
cases the error in the spatial measurement of the true reflector position
could be of the order of a kilometre~\citep{yilmaz}. Naturally, when it comes
to exploring hydrocarbon reserves, this lack of
precision will translate into a huge and wasted monetary expense. So an 
accurate representation of the depth and the local dip of subsurface reflectors
is imperative. Migration of seismic data goes a long way in resolving this
most difficult question~\citep{lowrie, yilmaz}. 

It should be useful for any study on the relevance of migration to start 
with a simple and pedagogic understanding of reflection 
seismology~\citep{lowrie, mgc}, whose primary purpose is to find the 
depths to reflecting interfaces and the corresponding seismic velocities 
of subsurface rock layers.  
This process first necessitates the artificial generation of mechanical 
waves at predetermined times
and spatial positions on the surface of the earth. These waves pass through
various earth layers and are reflected from the subsurface interfaces  
back to the surface of the earth, where they are recorded by receivers. 
For immediate purposes, the waves of interest are compressional (longitudinal)
waves, which are also referred to as primary waves~\citep{lowrie, mgc}.  

\begin{figure}[t]
\begin{center}
\includegraphics[angle=0,scale=0.6]{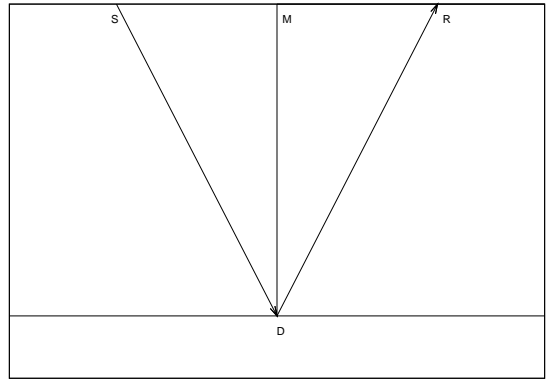}
\caption{\small{A schematic representation of reflection seismics from a 
flat subsurface stratum, using a single source-receiver pair. 
The waves originate from the source position, $\mrm S$. They are reflected 
from an absolutely flat interface in the subsurface,
back to the receiver position, $\mrm R$. The midpoint between $\mrm S$ and
$\mrm R$, shown as $\mrm M$, is vertically above the point from where the 
reflection occurs, indicated as $\mrm D$.
For many source-receiver arrays placed as symmetrically about $\mrm M$,
as it has been shown in the figure, the reflection point $\mrm D$ would 
be a {\em common} depth point for all such source-receiver pairs. The 
source-to-receiver distance, $\overline{\mrm{SR}}$, is $x$. The depth to
the reflecting interface, $\overline{\mrm{MD}}$, is given as $d$. The signal
takes a time $t$, travelling with a constant velocity $v$, to reach $\mrm R$
from $\mrm S$, after undergoing a reflection at $\mrm D$. Triangle 
$\overline{\mrm{SDR}}$ is an isosceles triangle, while triangle 
$\overline{\mrm{MDR}}$ is a right-angled triangle.}} 
\label{f1}
\end{center}
\end{figure}

Considering the simplest possible case of this system, one might suppose 
the surface of the earth to be absolutely flat. From this surface a single
source is creating waves which travel through the subsurface and are 
subsequently reflected back to the surface from a single stratum that is 
also equally flat. This has been schematically represented in Fig.~\ref{f1}.
In this simple situation
it is possible to set down the geometrical locus for the physical raypath 
in the $x$--$t$ plane, with $x$ defining the source-to-receiver position, 
and with the receiver itself gathering reflected signals after an interval 
of time, $t$.  

\begin{figure}[t]
\begin{center}
\includegraphics[angle=0,scale=0.6]{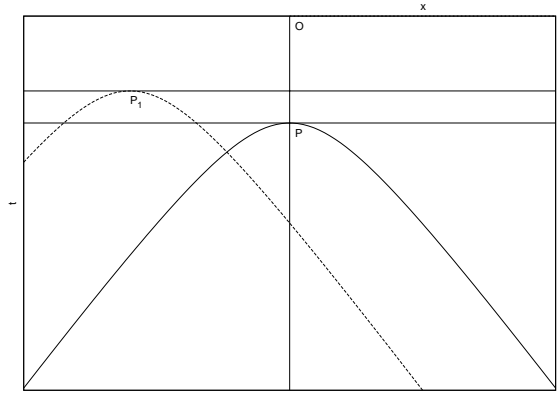}
\caption{\small{The locus of the raypath between a source and a receiver
in the $x$--$t$ plane. In this plot, $t$ increases in the downward direction
along the vertical axis. The origin of coordinates, $(0,0)$, is at $\mrm O$. 
The continuous curve corresponds to the source-receiver array shown in 
Fig.~\ref{f1}. The peak of this curve is at the point $\mrm P$, whose 
coordinates are $(0,t_0)$ in this $x$--$t$ plot. The dotted curve corresponds
to the situation shown in Fig.~\ref{f2}. The two horizontal lines are the 
tangents passing through the maxima of the two hyperbolae. 
It is evident from this plot that
inclined subsurface reflectors cause a shift in the locus of the raypath. In 
this case the maximum point on the curve has shifted to the point $\mrm P_1$.
With continuous change in the dip angle of the subsurface reflector, the path 
traced out by the peaks of the hyperbolae is an ellipse.}} 
\label{f15}
\end{center}
\end{figure}

To that end the Pythagoras theorem may 
be applied to the triangle $\overline{\mrm{MDR}}$ 
in Fig.~\ref{f1}, under the prescription that $\overline{\mrm{MR}} = x/2$,
$\overline{\mrm{MD}} = d$ (the vertical depth of the reflecting interface from 
the surface of the earth), and $\overline{\mrm{DR}} = vt/2$ (with $v$ being 
the constant velocity of the waves propagating through the subsurface). Followed 
by some simple algebraic manipulations, this exercise will deliver the $x$--$t$ 
locus corresponding to the actual raypath, known also as the {\em{normal moveout 
equation}}~\citep{lowrie}, as  
\begin{equation}
\label{hyper1}
\left(\frac{t}{t_0}\right)^2 - \left(\frac{x}{vt_0}\right)^2 = 1 ,
\end{equation}
which is an expression that very recognisably bears the canonical form of 
a hyperbola, with the parameter $t_0$ having  
been defined as $t_0 = 2d/v$. The solution given by Eq.~(\ref{hyper1}) 
has been geometrically depicted by the continuous curve in Fig.~\ref{f15}. 
From the plot
it may also be appreciated that different values of $d$, as a parameter in
Eq.~(\ref{hyper1}), will yield a family of hyperbolae (assuming that $v$
retains the same constant value all through). 
The turning points, the maxima in this case, of this family of 
hyperbolae (represented by the single continuous curve in Fig.~\ref{f15}),
obtained under the condition ${\mrm d}t/{\mrm d}x = 0$, will correspond to
the coordinate $(0, t_0)$ in the $x$--$t$ plane.
And for any fixed value of $x$ 
in the $x$--$t$ plane, the entire range of points along the $t$ axis (with 
$t$ customarily having been scaled along the vertical direction), cutting 
down the entire family of hyperbolae, will define 
what is known in seismic processing terms as a {\em seismic trace}~\citep{mgc}.

Recast in a slightly different form, Eq.~(\ref{hyper1}) will read as 
\begin{equation}
\label{hyprecast}
t = t_0 \sqrt{1 + \frac{x^2}{4 d^2}} , 
\end{equation} 
in which $t_0$ can physically be seen to represent the ``echo time" --- the 
two-way travel time for the 
wave, corresponding to a vertical reflection from the subsurface reflector. 
The term under the square root sign is the {\em normal moveout} factor,
which appears because the wave reaching the receiver at a distance $x$
from the shot point (i.e. the location of the source), has not been reflected 
vertically~\citep{lowrie}.
Under the condition that the receiver distance, $x$, is much less than 
the reflector depth, $d$, i.e. $x \ll d$, it is a matter of some simple 
algebraic steps to arrive at an approximate solution (through a binomial 
expansion) for the normal moveout, $\Delta t = t - t_0$, as
\begin{equation}
\label{nmo}
\Delta t \simeq \frac{x^2}{2 v^2 t_0} . 
\end{equation}  

The primary objective here is to determine the depth, $d$, of the reflector. 
The vertical echo time, $t_0$, and the normal moveout time, $\Delta t$, can
be obtained from the reflection data. The corresponding receiver distance, $x$, 
will also be known from the data. Taken together, all of these will allow 
for determining the value of $v$ from Eq.~(\ref{nmo}). 
This, with the help of the definition of $t_0$, will subsequently also allow 
for the determination of $d$. In all of this there is a clear signal 
that a precise extraction of the
value of $v$ from the reflection data is a principal necessity in determining
the reflector depth accurately. Even the simplest possible case that is being
studied here emphasises this fact. Knowing the precise value of the velocity 
will be an equally important issue when one considers more complicated instances
of reflection seismology involving non-horizontal reflectors of continually 
varying gradients. Besides this, with $v$ being dependent on the elastic
properties of the material through which the wave propagates~\citep{ll, mgc},
getting a correct measure of the velocity also conveys an impression of the
true nature of the subsurface material. 

\section{Dipping reflectors: A case for migration}
\label{sec2}

It is scarcely to be expected that the layering of actual geological strata will
conform to the neat and orderly picture implied by Fig.~\ref{f1}. If anything, not only
will the subsurface strata be quite great in number (instead of the single stratum
shown in Fig.~\ref{f1}), but also each stratum will have complex gradients in 
space. This will cause for much difficulty in devising an accurate theoretical 
model for the physics of reflection seismics. This fact can be amply illustrated 
even by considering the very simple case of a single dipping interface with a constant 
spatial gradient. 

\begin{figure}[t]
\begin{center}
\includegraphics[angle=0,scale=0.6]{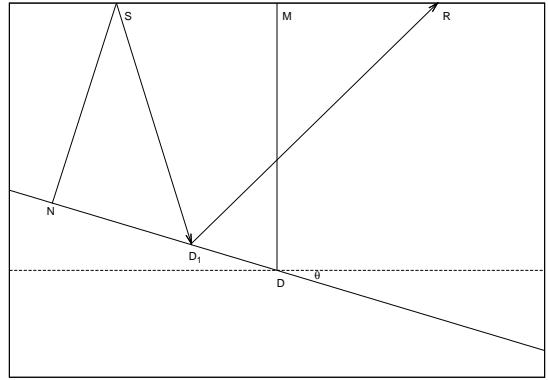}
\caption{\small{A schematic representation of reflection seismics from an inclined
subsurface reflector, using a single source-receiver pair placed symmetrically
about the midpoint, $\mrm M$. The inclined reflector is at a dip angle $\theta$
with respect to the horizontal. The actual point of subsurface reflection,
$\mrm{D_1}$, is {\em not} vertically below $\mrm M$ (unlike the point $\mrm D$),
and its location will vary for each set of source-receiver pair placed symmetrically
about $\mrm M$. There will be no common depth point for these multiple source-receiver 
pairs, because of the asymmetry arising from the dip. The perpendicular distance from 
the source to the reflector, $\overline{\mrm{SN}} = d$. The total travel path for the 
wave is 
$\overline{\mrm{SD_1}} + \overline{\mrm{D_1 R}} = vt$, and the source-to-receiver
distance, $\overline{\mrm{SR}} = x$.}}
\label{f2}
\end{center}
\end{figure}

When the reflecting interface is inclined at a constant angle $\theta$, with respect 
to the horizontal, as it has been shown in Fig.~\ref{f2}, it will be necessary to 
make suitable modifications in the analysis presented in Section~\ref{sec1}, to 
retain the canonical form of the hyperbolic normal moveout equation
that Eq.~(\ref{hyper1}) represents. This can indeed be achieved by noting in 
Fig.~\ref{f2} that the perpendicular distance from the source to the reflector,
$\overline{\mrm{SN}} = d$, the source-to-receiver distance,
$\overline{\mrm{SR}} = x$, and the total travel path, 
$\overline{\mrm{SD_1}} + \overline{\mrm{D_1 R}} = vt$. 
With this information, even upon 
accounting for the dip in the reflector, the invariant canonical 
form of the hyperbola can be obtained as 
\begin{equation}
\label{hyper2}
\left(\frac{t}{\bar{t}_0}\right)^2 - 
\left(\frac{\bar{x}}{v \bar{t}_0}\right)^2 = 1 , 
\end{equation} 
with $\bar{x}$ and $\bar{t}_0$ being defined by the transformations,  
$\bar{x} = x - 2 d \sin \theta$
and $\bar{t}_0 =(2d\cos \theta)/{v}$, respectively. 

From Eq.~(\ref{hyper2}), the coordinates
of the maximum point of the hyperbola $(x_{\mrm m},t_{\mrm m})$ in the $x$--$t$ 
plane (with $t$ increasing in the downward direction) will be given by the condition 
${\mrm d}t/{\mrm d}x = 0$. This will place the peak of the hyperbola at 
$(-2 d \sin \theta , \bar{t}_0)$. Going back to the case of the horizontal
reflector $(\theta = 0)$, given by Eq.~(\ref{hyper1}), the peak of the
hyperbola here is to be seen at $(0, t_0)$. Therefore, a distinct shift is seen to 
arise because of the dip of the subsurface reflector, and this shift has been 
shown by the position of the dotted hyperbola in Fig.~\ref{f15}. 

The locus of this shift can also be traced on the 
$x$--$t$ plane. Noting that the coordinates of the maximum of the normal moveout 
hyperbola are given by $x_{\mrm m}=-2d\sin \theta$ and $t_{\mrm m}=\bar{t}_0$, 
one could, by making use of the standard trigonometric result, 
$\sin^2 \theta + \cos^2 \theta = 1$, derive the condition, 
\begin{equation}
\label{locusmax}
\left(\frac{x_{\mrm m}}{2d}\right)^2+\left(\frac{t_{\mrm m}}{2d/v}\right)^2=1 .
\end{equation}
The form of this conic section is that of an ellipse. From this it is now possible
to conclude that the peak of the normal moveout curve will trace an elliptical
trajectory in the $x$--$t$ plane, with a continuous change in the dip angle, $\theta$. 

\begin{figure}[t]
\begin{center}
\includegraphics[angle=0,scale=0.6]{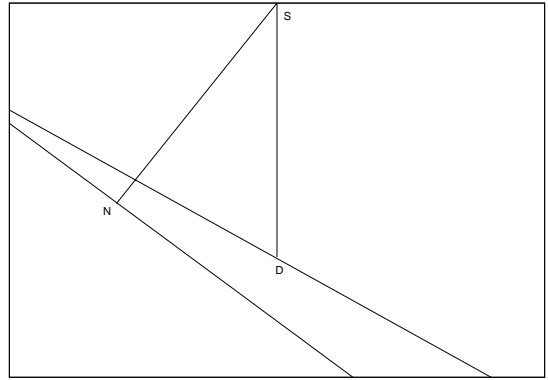}
\caption{\small{For an inclined reflector, imaging the depth vertically from 
the surface of the earth, gives a false position of the reflector. The actual 
point of reflection on the inclined plane is indicated by $\mrm N$, but a 
vertical downward projection from $\mrm S$ will give an erroneous impression 
that the reflection point is at $\mrm D$. The line $\overline{\mrm{SN}}$ is
normal to the reflector interface. Migration will give a more correct impression 
of the true reflector interface.}}
\label{f3}
\end{center}
\end{figure}

While this entire algebraic exercise looks very elegant in principle, it fails
hopelessly in addressing a serious practical problem. 
The derivation of Eq.~(\ref{hyper2})
has been done with the help of a single source-receiver pair. It {\em assumes} 
that every source-receiver pair in a full array of sources and receivers will have 
a {\em common} point of reflection from the inclined subsurface. {\it A priori}
this is not possible to achieve in practice. If anything, for a realistic arrangement
of source-receiver arrays, {\em each source-receiver pair
will have its own particular reflection point}. Therefore, unlike the simple
representation in Fig.~\ref{f1}, there will be no common depth point vertically
below the common midpoint of the source receiver arrays. Being unmindful of this
crucial point will only serve to furnish an erroneous impression of the subsurface
depth, of the kind that has been schematically shown in Fig.~\ref{f3}.
This fact categorically 
precludes the simple approach of determining the value of $d$, using the arguments 
which have been presented following Eq.~(\ref{nmo}), at the end of Section~\ref{sec1}. 

The difficulties do not end here. In actual fact the situation is vastly more
complicated. The dipping reflectors will have a varying gradient from one spatial
position to another, as opposed to being at a constant incline with respect to the
horizontal, shown in Fig.~\ref{f2}. Besides this, there will be geological faults,
and the propagation velocity of the seismic waves will also have a variation 
with respect to the depth. All of these facts will combine to render the accurate 
imaging of the subsurface a quite formidable task. Nevertheless, all reflection 
seismic records must be corrected for non-vertical reflections. This evidently 
complicated process of correction is what is, for all practical purposes,
implied by the term {\em migration}. It purports to shift all dipping reflections
to their true subsurface positions~\citep{lowrie, yilmaz}.  

Another important point has to be stressed here. So far the entire discussion 
has accounted for dip and subsurface variations in the {\em inline} 
direction~\citep{yilmaz}, 
i.e. the direction along which the source-receiver arrays have been lined up. 
This means that the analytical treatment presented so far has been confined to 
a $2D$ plane only --- along the vertical depth (given by the $z$ coordinate) and 
along the offset direction (given by the $x$ coordinate). However, practically 
speaking, variations in the subsurface will also take place in the {\em crossline} 
direction~\citep{yilmaz}, i.e. along the $y$ axis. So a comprehensive migration 
process must 
account for all variations along all the three spatial coordinates, i.e. it should 
in effect be a complete $3D$ process accounting for any change in the gradient 
of any subsurface reflector. 

\section{Migration: A mathematical theory}
\label{sec3}

By now the case for a complete $3D$ migration process has been fully argued for. 
To carry out $3D$ migration effectively, a necessary exercise would be to study  
the propagation of a compressional wavefield, $P \equiv P(x,y,z,t)$. This is 
described mathematically by the wave equation
\begin{equation}
\label{laplace}
\nabla^2 P - \frac{1}{v^2}\frac{\prt^2 P}{\prt t^2} = 0 , 
\end{equation}
with $\nabla^2$ being the standard Laplacian operator~\citep{aw}, and 
with $v$ having 
a general dependence on all the three spatial coordinates. Expressed in terms 
of its individual components, this second-order partial differential equation 
in both the space and the time coordinates will read as
\begin{equation}
\label{wave}
\left(\frac{\prt^2}{\prt x^2} + \frac{\prt^2}{\prt y^2} + \frac{\prt^2}{\prt z^2}
- \frac{1}{v^2}\frac{\prt^2}{\prt t^2} \right)P = 0 . 
\end{equation} 

On the surface of the earth the upcoming seismic wavefield would be measured as
$P(x,y,0,t)$. For fixed values of $x$ and $y$, the signal would be a function of
$t$ only, giving a single trace. Extended over the whole $x$--$y$ plane, 
this will build a proper mathematical understanding of the entire
$3D$ volume of seismic traces that one is required to work with. Given this 
{\em boundary} condition on the surface of the earth, one would have to find the 
wavefield, $P(x,y,z,0)$, which is actually the reflectivity of the earth in its 
recesses (and effectively, it also defines an {\em initial} condition at $t=0$). 
This whole process will require, as an 
intermediate step, the projecting of the surface wavefield to a depth $z$, 
corresponding to $t=0$. {\em In mathematical terms, this is the most general 
understanding that one can have about migration}~\citep{yilmaz}. 

Solving Eq.~(\ref{wave}) is no easy feat, because of its dependence on four
independent variables. However, under the simple working assumption that $v$
is a constant, it is possible to go a long way, by adopting the method of Fourier
transforms~\citep{aw}. By definition, the Fourier transform of the wavefield
$P(x,y,z,t)$ over the coordinates $x$, $y$ and $t$ is given as~\citep{aw}
\begin{equation}
\label{fou}
\tilde{P}(k_x, k_y, z, \omega) = \frac{1}{(2 \pi)^{3/2}} \int \int \int 
P(x,y,z,t) \exp \left(i k_x x + i k_y y - i \omega t \right) {\mrm d}x \,
{\mrm d}y \, {\mrm d}t . 
\end{equation} 
In terms of this transformed wavefield, Eq.~(\ref{wave}) can be recast as a
second-order ordinary differential equation, which is 
\begin{equation}
\label{orddiff}
\frac{{\mrm d}^2 \tilde{P}}{{\mrm d} z^2} 
+ \left(\frac{\omega^2}{v^2} - k_x^2 - k_y^2 \right)\tilde{P} = 0 , 
\end{equation} 
with $\tilde{P}$ now being understood to 
be $\tilde{P} \equiv \tilde{P}(k_x, k_y, z, \omega)$. 

It is worth stressing the importance of the method that has led 
to Eq.~(\ref{orddiff}). The starting point was Eq.~(\ref{laplace}) ---
a second-order partial differential equation in four variables. The Fourier transform
of $x$, $y$ and $t$, into their corresponding reciprocal domains of $k_x$, $k_y$
and $\omega$, has rendered this partial differential equation into a 
relatively simple 
second-order ordinary differential equation. For mathematical ease, this is 
the most important purpose that a Fourier transform can serve. Of course,
working in the Fourier domain has other practical advantages, especially from
the perspective of the processing geophysicist, like the 
elimination of spikes and noise from the data~\citep{mgc}. 

Following the Fourier transform, it will now be 
necessary to solve Eq.~(\ref{orddiff}) 
to extrapolate the wavefield from its value on the surface of the earth (at $z=0$)
to the required depth (where $z \neq 0$). Working under the simplifying assumption 
that $v$ is constant~\citep{yilmaz}, it becomes quite easy to solve Eq.~(\ref{orddiff}), 
which will actually have the form of the simple harmonic oscillator~\citep{aw}. 
With the imposition of the definition~\citep{yilmaz} 
\begin{equation}
\label{kayzed}
k_z^2 = \frac{\omega^2}{v^2} - k_x^2 - k_y^2 
\end{equation} 
(a {\it dispersion relation} as it were\footnote{In the case of the so-called
{\em Exploding Reflectors} model, the velocity in Eq.~(\ref{kayzed}) is 
transformed according to the rule, $v \longrightarrow v/2$~\citep{yilmaz}.}), 
one can proceed to derive an integral solution of Eq.~(\ref{orddiff}) as
\begin{equation}
\label{solinteg}
\tilde{P}(k_x, k_y, z, \omega) = 
\tilde{P}(k_x, k_y, 0, \omega) \exp \left(- i k_z z \right) . 
\end{equation} 
Actually, the most general form of Eq.~(\ref{solinteg}) is given by the 
superposition of an upcoming wave and a downgoing wave. i.e. 
$\exp ( \pm i k_z z )$. However, only the solution with the negative
sign need be retained here, since it corresponds physically to upcoming waves, 
which, for the present purpose, are actually the relevant modes~\citep{yilmaz}. 
This whole analysis shows how it should be possible  
to extrapolate the surface wavefield
to a given depth below the surface of the earth. While it has been quite 
a simple task to arrive 
at this result under the assumption of constant velocity, one can also derive 
an asymptotic solution for the extrapolation exercise when the velocity is 
dependent on the depth, i.e. $v \equiv v(z)$. This treatment has been 
presented in Section~\ref{sec35}. 

One way or the other, the one primary fact that stands out is that 
so far as migration is concerned,
the result given by Eq.~(\ref{solinteg}), i.e. extrapolating the surface data to
the depth, represents a crucial step. Beyond this point there can be many variations
in the subsequent execution of the migration process, depending on varying contingencies.
In what follows later, the specific principles of some of these methods will be 
dwelt upon.

\section{Extrapolating the wavefield in the case of depth-dependent velocity: The
WKB approximation}
\label{sec35}

It has already been seen how easily Eq.~(\ref{orddiff}) can be integrated when $v$ 
is constant. Supposing now that $v$ has a dependence on the $z$ coordinate, i.e.
$v \equiv v(z)$, it will still be possible to derive an asymptotic solution 
of Eq.~(\ref{orddiff}),
under appropriate conditions, with the help of the {\rm WKB}
({\it Wentzel-Kramers-Brillouin}) approximation~\citep{mw, kbo}.

Since $v$ has a dependence on $z$, so should $k_z$ have a dependence on $z$, by 
virtue of Eq.~(\ref{kayzed}). Therefore, one can write
\begin{equation}
\label{a1}
\frac{{\mrm d}^2 \tilde{P}}{{\mrm d} z^2}
+ k_z^2(z) \tilde{P} = 0 ,
\end{equation}
for which a trial solution in the most general form can be set down as 
\begin{equation}
\label{a2}
\tilde{P} = A(z) e^{i \alpha (z)} , 
\end{equation} 
with $A$ being an amplitude function, and $\alpha$ being a phase function. 
Substituting this trial solution in Eq.~(\ref{a1}) will give
\begin{equation}
\label{a3} 
\left [ \frac{1}{A} \frac{{\mrm d}^2 A}{{\mrm d}z^2} - 
\left(\frac{{\mrm d} \alpha}{{\mrm d} z} \right)^2 + k_z^2(z) \right ] 
+ i \frac{{\mrm d} \alpha}{{\mrm d} z} \frac{\mrm d}{{\mrm d}z} 
\left[ \ln \left(\frac{{\mrm d} \alpha}{{\mrm d} z} A^2 \right) \right] = 0 . 
\end{equation} 
The real and imaginary components above will have to be set to zero separately.
In doing this, a solution for the imaginary part can be extracted, and it will 
read as 
\begin{equation}
\label{a4}
A = C \bigg{\vert} \frac{{\mrm d} \alpha}{{\mrm d} z}\bigg{\vert}^{-1/2} ,
\end{equation}
where $C$ is a constant of integration. The foregoing result indicates that 
knowing either $A$ or $\alpha$ independently will suffice to give the full 
solution implied by Eq.~(\ref{a2}).  

So far the mathematical treatment has been exact. 
In deriving a solution from the real component of Eq.~(\ref{a3}), it will be
necessary now to apply the
{\rm WKB} approximation, requiring that the amplitude function $A(z)$ 
in Eq.~(\ref{a2}), is a much more 
slowly varying function of $z$ than the phase, $\alpha (z)$. 
In that event one can neglect the second derivative of $A$ with respect 
to $z$ in Eq.~(\ref{a3}). This will give the phase solution 
\begin{equation}
\label{a5}
\alpha (z) \simeq - \int k_z(z) \, {\mrm d} z ,
\end{equation} 
(taking the more practically relevant negative solution, as usual) 
which, used along with Eq.~(\ref{a4}), will give an aysmptotic solution 
that will read as 
\begin{equation}
\label{a6}
\tilde{P} \simeq \frac{C}{\sqrt{k_z}} \exp \left[ - i \int k_z(z) \, {\mrm d} z \right] . 
\end{equation} 

How can the use of the {\rm WKB} approximation be justified self-consistently? On
using the solutions given by Eqs.~(\ref{a4}) and~(\ref{a5}), it can be seen that
the second derivative of $A$ is given as 
\begin{equation}
\label{a7}
\frac{{\mrm d}^2 A}{{\mrm d}z^2} \simeq A \left(\frac{\omega}{k_z v}\right)^2 
\Bigg\{  \left[2 - \frac{5}{2} \left(\frac{\omega}{k_z v}\right)^2 \right]
\times \left[ \frac{\mrm d}{{\mrm d}z} \left(\ln v \right) \right]^2 - 
\frac{\mrm d^2}{{\mrm d}z^2} \left(\ln v \right) \Bigg\} . 
\end{equation} 
On the right hand side of Eq.~(\ref{a7}) one can see derivatives of the logarithm 
of $v$. Now it is known that the logarithm of any function varies much more slowly 
than that function itself. The same should hold for $\ln v$ in comparison with $v$ 
itself. Therefore, it becomes eminently possible to argue that with the second 
derivative of $A$ depending on the derivatives of $\ln v$, 
the requirement of $A$ being a slowly varying function of $z$ is satisfactorily
fulfilled, and one may safely neglect the term
containing the second derivative of $A$ in Eq.~(\ref{a3}). Indeed, this will be all
the more appropriate for greater depths, where, with $v$ being an increasing function
of $z$, the logarithmic terms will strongly temper the variation of $A$ arising due 
to the velocity variations. So the asymptotic solution given by Eq.~(\ref{a6}) will 
have a firmer validity at greater depths. This entire 
line of reasoning upholds the invoking 
of the {\rm WKB} approximation. 

\section{The $3D$ phase-shift migration in the frequency-wavenumber domain}
\label{sec4}

It has been discussed in Section~\ref{sec3} that the very definition of migration 
necessitates projecting seismic data from the {\em boundary} condition $z=0$ for 
all $t$, to the {\em initial} condition $t=0$ for all $z$. 
To this end, it shall be expedient to set down first the inverse Fourier
transform corresponding to Eq.~(\ref{fou}). This will be 
\begin{equation} 
\label{invfou}
P(x, y, z, t) = \frac{1}{(2 \pi)^{3/2}} \int \int \int
\tilde{P}(k_x, k_y, z, \omega) 
\exp \left(- i k_x x - i k_y y + i \omega t \right){\mrm d}k_x \,
{\mrm d}k_y \, {\mrm d}\omega .
\end{equation}
It is also known by now how it should be possible to 
relate $P(k_x, k_y, z, \omega)$ to the 
surface seismic data with the help of Eq.~(\ref{solinteg}). Using this condition 
in Eq.~(\ref{invfou}) and setting $t=0$, will give
\begin{equation}
\label{imagt0}
P(x, y, z, 0) = \frac{1}{(2 \pi)^{3/2}} \int \int \int
\tilde{P}(k_x, k_y, 0, \omega) 
\exp \left(- i k_x x - i k_y y - i k_z z \right){\mrm d}k_x \,
{\mrm d}k_y \, {\mrm d}\omega , 
\end{equation} 
which is the equation for the phase-shift method, suited for $3D$ migration~\citep{yilmaz}. 
The involvement of $\omega$ can be eliminated from this equation, by making use of
Eq.~(\ref{kayzed}), under the constraint that $v$ is constant, and $k_x$ and $k_y$ are 
to be kept unchanged~\citep{yilmaz}. This will first deliver the relation
\begin{equation}
\label{omega}
{\mrm d} \omega = \frac{v k_z}{\sqrt{k_x^2 + k_y^2 + k_z^2}} {\mrm d} k_z , 
\end{equation}
leading ultimately to 
\begin{displaymath}
P(x, y, z, 0) = \frac{1}{(2 \pi)^{3/2}} \int \int \int 
\frac{v k_z}{\sqrt{k_x^2 + k_y^2 + k_z^2}} 
\tilde{P} \left(k_x, k_y, 0, v \sqrt{k_x^2 + k_y^2 + k_z^2} \right)
\end{displaymath}
\begin{equation}
\label{stolt} 
\qquad \qquad \qquad \qquad \qquad \qquad \qquad \qquad 
\times
\exp \left(- i k_x x - i k_y y - i k_z z \right){\mrm d}k_x \,
{\mrm d}k_y \, {\mrm d}k_z , 
\end{equation} 
which is the equation for constant velocity $3D$ Stolt migration~\citep{yilmaz}. 

The derivation of the result in Eq.~(\ref{stolt}) is certainly consistent with 
what has been forwarded as a mathematical definition of migration in Section~\ref{sec3}. 
However, short of resorting to an involved numerical treatment, the 
full $3D$ problem does not lend itself very easily to deriving any pedagogical
insight out of it. Hence it should be instructive to consider a very special case 
of this migration method, for a wavefield that has a spatial dependence on the 
$z$ coordinate only, and so will actually be a vertically propagating planar wavefield. 

For such a wavefield, with no dependence on the $x$ and $y$ coordinates,
and propagating in a planar front along the $z$ axis only, one can
set down the wavefield as $P \equiv P(z,t)$. In terms of the Fourier inverse
transform it can be expressed as
\begin{equation}
\label{b1}
P(z,t) = \frac{1}{(2 \pi)^{1/2}} \int \tilde{P} (z, \omega) \exp
\left( -i \omega t \right) {\mrm d} \omega .
\end{equation}
Now it is already known that Eq.~(\ref{solinteg}) will give an extrapolation
relation going as
\begin{equation}
\label{b2}
\tilde{P}(z, \omega) =
\tilde{P}(0, \omega) \exp \left(- i k_z z \right) ,
\end{equation}
for the special case that is being studied here. From Eq.~(\ref{kayzed}) one
also gets $k_z = \omega/v$, since there is no involvement of the $x$ and $y$
coordinates. Use of all these conditions in Eq.~(\ref{b1}), will result in
\begin{equation}
\label{b3}
P(z,t) = \frac{1}{(2 \pi)^{1/2}} \int \tilde{P} (z, \omega) \exp
\left[ -i \omega \left( \frac{z}{v} + t \right)\right] {\mrm d} \omega ,
\end{equation}
which, for $t=0$, will give
\begin{equation}
\label{b4}
P(z,0) = \frac{1}{(2 \pi)^{1/2}} \int \tilde{P} (0, \omega) \exp
\left[ -i \omega \left( \frac{z}{v} \right)\right] {\mrm d} \omega =
P(0, z/v) .
\end{equation}
But it is also known that $t$ and $\omega$ are conjugate variables for the
Fourier transform. Their mathematical connection is given by
\begin{equation}
\label{b5}
P(0,t) = \frac{1}{(2 \pi)^{1/2}} \int \tilde{P} (0, \omega) \exp
\left( -i \omega t \right) {\mrm d} \omega ,
\end{equation}
which can then be compared with the result given in Eq.~(\ref{b4}). This will
immediately allow for setting down
\begin{equation}
\label{b6}
z = vt ,
\end{equation}
which is a result that very unambiguously connects the depth of a reflecting 
stratum with the time of the reception of a signal reflected from that stratum  
(under the simple
condition that the velocity of the wave propagation is unchanging). An important
point to note here is that the simple appearance of Eq.~(\ref{b6}) has been 
possible because the plane wave allows only the vertically propagating mode
in Eq.~(\ref{kayzed}), and so by default this mode has 
been decoupled from all the other modes. This decoupling should be impossible 
to achieve for higher spatial dimensions, as a look at Eq.~(\ref{kayzed}) ought
to reveal. 

\section{The stationary phase approximation}
\label{sec5}

It shall be useful at this point to derive the traveltime equation for the $3D$ 
zero-offset
case, using the method of the stationary phase approximation. From Eq.~(\ref{invfou})
one can, making use of the extrapolation condition given by Eq.~(\ref{solinteg}),
write a compact relation that will read as 
\begin{equation}
\label{statio}
P(x, y, z, t) = \frac{1}{(2 \pi)^{3/2}} \int \int \int
\tilde{P}(k_x, k_y, 0, \omega)
\exp \left( i\Phi z \right){\mrm d}k_x \,
{\mrm d}k_y \, {\mrm d}\omega , 
\end{equation} 
in which 
\begin{equation}
\label{phi}
\Phi = - k_z - \frac{k_x x}{z} - \frac{k_y y}{z} + \frac{\omega t}{z} . 
\end{equation} 

It is easy to see that the integrand in Eq.~(\ref{statio}) has an amplitude part
and a phase part. If the phase is to vary much more rapidly than the amplitude
(a requirement that is consistent with the {\rm WKB} analysis presented in 
Section~\ref{sec35}), then the consequent rapid oscillations over most of the range 
of the integration, will result in an average value of nearly zero. The exception
to this principle will occur when $\Phi$ is stationary~\citep{mw, aw}. This (the 
stationary phase
approximation) will be quantified by $\delta \Phi = 0$, which is, in fact, a 
condition for the turning points of $\Phi$. The major contribution to
the integral will, therefore, come from the turning points of the phase function
$\Phi$. This condition will help in identifying the physical coordinates from where
one will obtain the most significant contribution to the signal received on the 
surface of the earth. 

Going by Eq.~(\ref{kayzed}), it is already known that $k_z$ has a dependence on 
$k_x$, $k_y$ and $\omega$. Taking this in conjunction with Eq.~(\ref{phi}), and 
noting that the relevant variables for the integration in Eq.~(\ref{statio}) 
are $k_x$, $k_y$ and $\omega$, the required extremum condition for $\Phi$ 
can consequently be stated as
\begin{equation}
\label{phivar}
\delta \Phi = \left(\frac{\prt \Phi}{\prt k_x}\right) \delta k_x
+ \left(\frac{\prt \Phi}{\prt k_y}\right) \delta k_y
+ \left(\frac{\prt \Phi}{\prt \omega}\right) \delta \omega = 0 . 
\end{equation} 
The result above can only mean that all the three partial derivatives in Eq.~(\ref{phivar}) 
are to be individually set to zero. This will give three distinct mathematical conditions,
from which one will be able to derive the traveltime equation for the $3D$ zero-offset
case as
\begin{equation}
\label{traveltime}
x^2 + y^2 + z^2 = \left(vt\right)^2 . 
\end{equation} 
This defines a sphere of radius $vt$ in the $x$--$y$--$z$ coordinate space, and a 
hyperboloid in the $x$--$y$--$t$ space for a constant value of $z$. Considering only
the inline direction and the depth (i.e. when $y=0$), Eq.~(\ref{traveltime}) will 
define a semi-circle below the surface of the earth in the $x$--$z$ plane, and the
usual hyperbola in the $x$--$t$ plane. 

However, a much more interesting result can be seen for the simplest possible case of the 
planar 
wavefront, i.e where the spatial dependence is only on the $z$ coordinate. In that case 
the result in Eq.~(\ref{traveltime}) will converge simply to the linear solution $z = vt$, 
which is exactly what Eq.~(\ref{b6}) also furnishes. So effectively this simple special 
case establishes an intriguing mathematical equivalence 
between the stationary phase approximation method and the constant velocity Stolt 
migration algorithm.

\section{Some general comments}
\label{sec6}

Migration strategies can be implemented over a wide range --- from $2D$ poststack 
time migration to $3D$ prestack depth migration. The geological features of the 
subsurface will determine the choice of a particular algorithm, or even a combination
of various algorithms. In practice, time migration enjoys wide popularity because it
is least sensitive to errors in velocity, and delivers acceptable results to make
a reliable interpretation. However, time migration is trustworthy only as long as
the lateral velocity variations are at most moderate. Otherwise depth migration is
more appropriate. 

Standard migration algorithms are based on the wave equation. One can classify 
migration processes under three broad categories~\citep{yilmaz}: 
\begin{itemize}
\item
Those based on the integral solution of the wave equation. 
\item
Those based on the finite-differencing of the partial differential
equation implied by the wave equation. 
\item
Those based on the implementation of the frequency-wavenumber algorithm. 
\end{itemize} 

It is to be borne in mind that none of the migration algorithms that 
are implemented in the industry today gives a complete prescription for handling
all dips and velocity variations, while being cost-effective at the same time. 
Migration algorithms based on the integral solution of the wave equation (Kirchhoff
migration) can become cumbersome when they encounter lateral velocity variations. 
Finite-difference algorithms can only be specified for different degrees of 
dip approximations. And frequency-wavenumber algorithms are effective against 
lateral velocity variations only to a limited extent. 

\section*{Acknowledgements}
%\acknowledgments
The author would like to thank J. K. Bhattacharjee, B. Bleines, S. Ghoshal, J. Kumar,
S. Maitra, S. Roy Chowdhury and S. Sarkar for their comments. The author also 
expresses his gratitude to A. Kumar and T. Mukherjee for providing some useful 
technical material, and to S. Hegde, P. Madkar, P. Pathak and J. Shaikh for 
help in various other relevant matters.

\end{document}